\documentclass{aa}

\usepackage{epsf}

\usepackage{times}
\usepackage{mathptm}

\begin{document}
\thesaurus{03                          
              ( 11.17.3,               
                11.19.1,               
                11.05.2,               
                11.12.2,               
                12.03.3                
                12.07.1                
               )}

\title{The bright end of the QSO luminosity function}

\author{Lutz Wisotzki }

\institute{Hamburger Sternwarte, Gojenbergsweg 112, D-21029 
           Hamburg, Germany,\protect\\
           e-mail: {\tt lwisotzki@hs.uni-hamburg.de}
           }
          
\date{Received ; accepted} 
\maketitle
\markboth{L. Wisotzki: Bright end of QSO luminosity function}{}

\begin{abstract}
We have analysed the optical luminosity-redshift distribution 
properties of bright QSOs, using a new large sample
from the Hamburg/ESO survey.
The sample provides insight into the hitherto poorly sampled bright 
tail of the luminosity function, allowing to monitor its evolution
with redshift up to $z\approx 3$.
The slope increases significantly towards higher $z$,
inconsistent with the predictions of
pure luminosity evolution, but also with other recently proposed 
parameterisations. This phenomenon is opposite to what would be
expected from gravitational lensing, showing that magnification
bias does not significantly distort the QSO luminosity function
within the redshift range covered.
The space density of high-luminosity QSOs continues to increase 
up to the high-redshift limit of the survey, 
without indication of reduced evolution above $z\simeq 2$.
The sample also permits an improved estimate of the local 
($z\approx 0$) luminosity function of QSOs and bright 
Seyfert~1 nuclei, over the luminosity range 
$-27\la M_{B_J}\la -20$. No evidence for a break or change
of slope is found down to absolute magnitudes $M_{B_J}\simeq-20$.
\end{abstract}

\begin{keywords}
Quasars: general -- Galaxies: Seyfert -- 
Galaxies: evolution -- Galaxies: luminosity function -- 
Cosmology: observations -- Gravitational lensing
\end{keywords}

\section{Introduction}

The wide range of quasar luminosities,
spanning altogether over four decades from low-activity Seyfert galaxies
up to the most luminous objects in the universe found at high $z$,
allows individual surveys to access only small portions of the
luminosity-redshift-plane at given redshift.
While at intermediate flux levels there are now many samples available, 
large uncertainties still exist at the faint and
bright tails of the QSO luminosity function (QLF).
The lack of good constraints for the bright end is particularly grave 
in the optical waveband, where the Palomar-Green
Bright Quasar Sample (BQS; Schmidt \& Green \cite{schm+gree:83:QEBQS})
has been the only significant contributor for now 15 years,
although ist is now well known to be substantially incomplete
(Goldschmidt et al.\ \cite{goldschmidt*:92:SDBQ};
K\"ohler et al.\ \cite{koehler*:97:LQLF}).

The most widely quoted analysis of the quasar luminosity function is
due to Boyle and collaborators 
(e.g., Boyle et al.\ \cite{boyle*:88:EOQ}).
who concluded that the evolution proceeds along increasing
luminosity, with the simple scheme of `pure luminosity evolution'
(PLE) providing a statistically acceptable fit to their data.
However, the Boyle et al.\ analysis had to rely on the BQS
and was therefore affected by the incompleteness of that sample.
Newer surveys sensitive to the brighter part of the QSO population
have questioned the validity of the PLE picture. In a preliminary 
analysis of the LBQS, Hewett et al.\ (\cite{hewett*:93:EBOQ}) 
showed that the PLE model predicts too few bright
and too many faint low-redshift quasars. A similar
conclusion was reached by Goldschmidt \& Miller 
(\cite{gold+mill:98:QE}) using the Edinburgh survey, by
La Franca \& Cristiani (\cite{lafr+cris:97:HBQS}) from a combined 
analysis of EQS and HBQS, and by K\"ohler et al.\
(\cite{koehler*:97:LQLF}) comparing their `local'
QLF with the PLE prediction.

For redshifts $z>2$, the general view is that the
evolution slows down considerably around $z\simeq 2$ and 
eventually reverses. The actual onset of this reversal is,
however, not well constrained, and may well depend on luminosity. 
In the PLE model of Boyle et al.\ (\cite{boyle*:91:QE}), comoving
space densities are roughly constant for redshifts $z\ga 1.9$,
while the results of Hewett et al.\ (\cite{hewett*:93:EBOQ})
and Warren et al.\ (\cite{warren*:94:MSQ3}) 
suggest that space densities still increase up to $z\simeq 3$,
although at a reduced rate, but drop precipitously thereafter. 
A marked decline in space densities towards higher $z$ was also 
found by Schmidt et al.\ (\cite{schmidt*:95:ELFQ}).
However, all these surveys sampled mainly intermediate-luminosity 
QSOs with $M_B \simeq -26$. 
For highest luminosities, $M_B\la -28$, there is very little
material available, and the few results so far are controversial.
Irwin et al.\ (\cite{irwin*:91:HZQ}) claimed that there is no 
evidence for evolution at all, up to $z=4.5$. This has been 
disputed by Kennefick et al.\ (\cite{kennefick*:95:LFQ}), 
who obtain an order of magnitude lower space densities at 
$z=4.3$ than Irwin et al. Resolving this controversy
will require new wide-angle survey material; the first
results from the just started Sloan Digital Sky Survey 
(Fan et al.\ \cite{fan*:99:HRQ}) indicate that data on 
high-luminosity QSOs in the very early universe will soon be 
available in substantial numbers.

In this paper we present the contribution of a new
large sample of bright QSOs, drawn from the Hamburg/ESO survey
(HES; Wisotzki et al.\ \cite{wisotzki*:96:HES1},
Reimers et al.\ \cite{reimers*:96:HES2}).
This sample contains many of the most luminous quasars known 
and is suited to study the evolutionary properties
at the bright part of the QLF over a wide range of redshifts. 
It is also the first quasar sample large enough to effectively
\emph{replace} the BQS, with higher completeness,
photometric accuracy, and less affected by redshift-dependent 
selection biases. The main aim of this paper is to use the HES
sample to derive constraints on the evolution of luminous QSOs.
We do not attempt to compute new evolutionary models, as that
would require merging the HES with other, fainter samples --
a complication that we wish to avoid in the present paper.

\begin{figure}[tb]
\epsfxsize=8.8cm
\epsfclipon
\epsfbox[60 76 456 371]{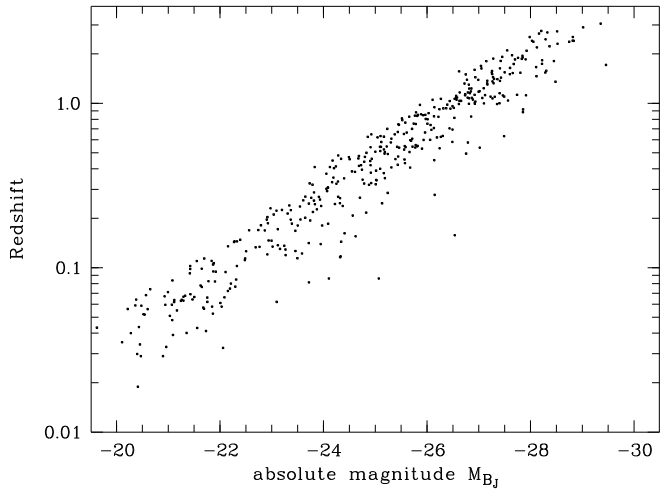} 
\caption[]{Distribution of the 415 QSOs of the flux-limited HES sample
over absolute magnitudes and redshifts.}
\label{fig:z_mabs}
\end{figure}

\section{Observational data}

The Hamburg/ESO survey was initiated in 1990 as an ESO key programme,
to perform a wide-angle search for bright QSOs in the southern sky. 
The survey uses objective prism plates taken with the ESO Schmidt
telescope and digitised with the Hamburg PDS microdensitometer
(for more details on the survey see 
Reimers \& Wisotzki \cite{reim+wiso:97:HES}; 
Wisotzki et al.\ \cite{wisotzki*:99:HES3} -- hereafter Paper~1). 

The selection of QSO candidates from the database of digital spectra
involves a multitude of selection criteria.
Extensive follow-up spectroscopy conducted at ESO has allowed to 
construct a new flux-limited sample of 415 objects within the redshift range
$0\la z < 3.2$, with optical magnitudes $B_J \la 17.5$;
a full discussion of the sample properties is given in Paper~1,
of which only some essentials are summarised here.
The sample was compiled from processing 207 HES Schmidt fields
with complete follow-up spectroscopic identification 
down to well-defined flux limits.
Each field has its own limiting magnitude depending on the
quality of photographic plates and seeing.
While one single Schmidt plate formally subtends over 
$\sim 5\degr\times 5\degr$, the \emph{effective survey area} 
is reduced because of overlapping adjacent plates and 
loss of processable area. The total effective area
$\Omega_{\mathrm{eff}}$ is 3700\,deg$^2$ for $B_J<14.5$,
and a monotonously falling function of magnitude for fainter objects.
The uncertainties of optical photometry are generally smaller than 0.2\,mag.
All redshifts were determined from
follow-up slit spectra, with typical continuum S/N ratios of $\ga 20$, and
for no object the assignment of an appropriate redshift was in doubt.

Due to its wide range of selection criteria and highly automated 
surveying procedure, the HES is less affected by
redshift-dependent selection biases than many other optical surveys.
In Paper~1 we give an extensive discussion of the completeness
for this sample; the central points are summarised as follows: \\
(1) The redshift distribution does not show evidence for significantly
enhanced or reduced selection efficiency in certain redshift regions.\\
(2) More than 99\,\% of the previously known QSOs located within the
survey area and above the flux limits were recovered. \\
(3) QSOs within the usual range of spectral properties are always selected. \\
(4) The surface density of bright QSOs as measured in the HES is
higher by a factor of 1.5 than in the BQS, and fully compatible
with other recent quasar surveys.

Given apparent magnitude $B_J$ and redshift $z$, absolute magnitudes
$M_{B_J}$ of the QSOs were estimated using the usual formula
$M_{B_J} = B_J + 5 - 5 \log d_L(z) + A + K(z)$,
where $d_L(z)$ is the `luminosity distance' in an expanding 
Friedmann universe, computed with the relation given by 
Terrell (\cite{terrell:77}). We adopted
$H_0 = 50\,\mathrm{km\,s}^{-1}\,\mathrm{Mpc}^{-1}$ and
$q_0 = 0.5$ to specify the cosmological model.
For the extinction term $A$,
only attenuation by Galactic dust has been assumed, and the values
of $A_{B_J}$ were estimated from H\,{\sc i} column densities
as described in Paper~1.

\begin{figure*}[tb]
\epsfxsize=12cm
\epsfclipon
\epsfbox[85 85 542 479]{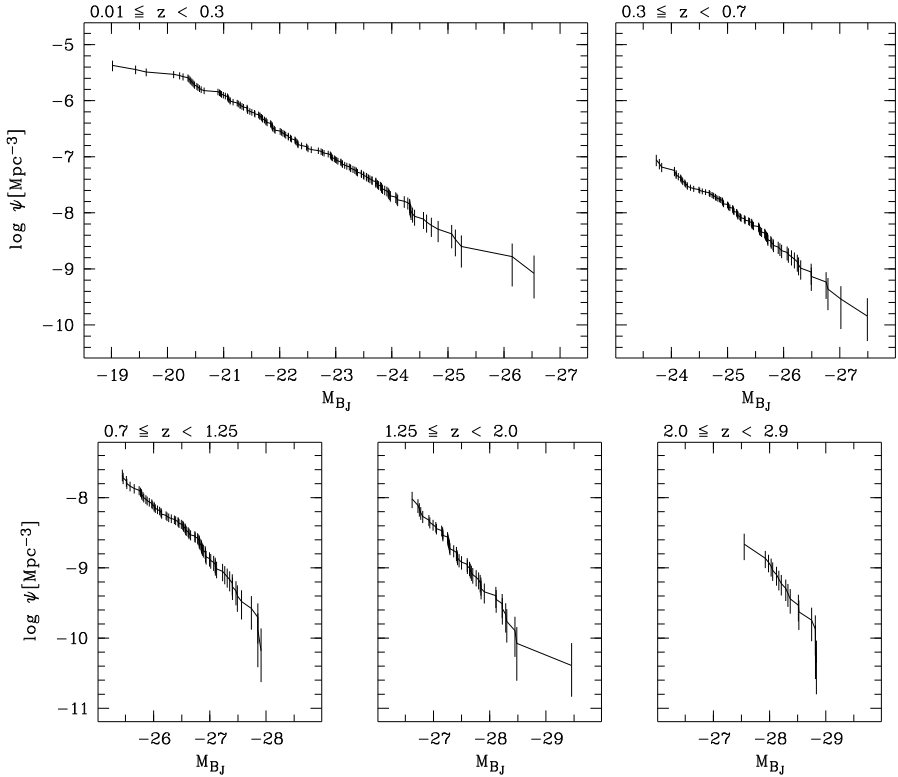} 
\hfill
\raisebox{2ex}{\parbox[b]{55mm}{
\caption[]{The cumulative quasar luminosity function
determined from the HES sample for five adjacent redshift shells.
Each panel shows the result for one redshift domain, 
including error bars estimated from Poisson statistics.}}}
\label{fig:cumlf_p}
\end{figure*}

Somewhat more complex is the situation for the $K$ term:
Although quasar spectra have often been approximated by simple power laws
$f_\nu \propto \nu^\alpha$ with adopted mean spectral index $\alpha$ 
between $-1$ and $-0.3$, actual spectral energy distributions 
are more complicated than this. We have recently determined a
new $K$ correction as a function of redshift that results in 
luminosities lower by 0.4\,mag for high-$z$ QSOs, compared to 
previously published relations (Wisotzki \cite{wisotzki:99:K}),
and we adopt that new $K(z)$ relation for the present paper.
The reader should be alert that the space densities derived in the
next section are therefore shifted towards lower luminosities,
relative to what can be found in the literature.
However, whenever comparisons with the results of earlier workers
are carried out in this paper, 
in particular with parametric descriptions of the QLF and its evolution,
we use the appropriate original $K(z)$ relations 
employed by others. The qualitative conclusions of this paper
do not depend on the details of the adopted $K$ correction.

Fig.\ \ref{fig:z_mabs} shows the
distribution of sources over absolute magnitudes and redshifts. 
At $z< 0.1$, the majority of objects would usually be called 
Seyfert~1 galaxies rather than QSOs,
but at all other redshifts there is no ambiguity of this sort,
the HES always sampling the optically brightest parts of the 
known quasar population.

\section{Luminosity functions} \label{sec:lf}

Space densities at given absolute magnitude were computed 
in bins of redshift, and luminosity for the binned differential form,
using the $1/V_{\mathrm{max}}$ estimator (Felten \cite{felten:76:ELF}).
Five redshift intervals were chosen to represent the evolution of the 
bright end of the QLF with cosmic time, with boundaries
$z = 0.0$--0.3, 0.3--0.7, 0.7-1.25, 1.25--2.0, and 2.0--2.9.
These intervals correspond to approximately equal steps in 
$\log (1+z)$ which is the cosmic time variable in the
PLE model of Boyle et al.\ (\cite{boyle*:91:QE}), 
from where the above set of intervals
was adopted. The lowest ($z=0$--0.3) interval is not common in
quasar evolution studies in the optical (see below), but has been
added for this analysis. 

Fig.\ \ref{fig:cumlf_p} displays the luminosity
function, in cumulative form, for each shell separately.
As discussed by Hewett et al. (\cite{hewett*:93:EBOQ}), this way
of representing a luminosity function has considerable advantages 
over the conventional approach to bin the data in absolute magnitude:
The contribution of each object is easily made apparent, the
complicated weighting within the luminosity bins needs not be corrected for,
and the full range of the data can be displayed without `incomplete bin' 
edge effects. However, the data are still binned in redshift and
the constructed luminosity functions therefore do not represent 
the QLF at specific epochs, but at `mean' epochs that vary with
luminosity (a form of Malmquist bias). In particular, in the presence 
of strong positive evolution, the apparent QLF for any given redshift shell
will appear considerably flatter than the intrinsic QLF. The statistical
test procedures employed below are not affected by this bias.
In the following we comment briefly on the observed properties of
each of the subsets. 

\noindent $\mathbf{z < 0.3}$: \ 
The local ($z \approx 0$) universe is the only domain where the full range of
luminosities is technically accessible to a single flux-limited survey.
The HES is the first optical quasar survey capable of fully exploiting this
option, being specifically designed to reduce selection biases due to
the presence of extended galaxy envelopes. This has already yielded a 
first estimate of the local QLF based on a small subsample of the HES
(K\"ohler et al.\ \cite{koehler*:97:LQLF}).
The present sample of 160 objects with $z<0.3$ increases 
the number of 20 `local QSOs' used by 
K\"ohler et al.\ by almost an order of magnitude, 
allowing a much more accurate assessment 
of QLF shape and slope. 
The local luminosity function is discussed further
in Sect.~\ref{sec:lqlf} below.

\noindent $\mathbf{0.3 < z < 0.7}$: \ 
This shell is equal to the `low-redshift' regime 
of most other optical quasar surveys. 
Host galaxies are visible only in exceptional cases, due to
the $(1+z)^4$ surface brightness dimming and the steep $K$ correction for
galaxy spectra in the $B_J$ band. 
For $z<0.5$, quasar spectra are generally extremely
blue and easily separated from their UV colours;
around $z\simeq 0.6$, the short-wavelength end of the `little blue bump'
causes quasar colours to be somewhat redder, but it is shown in
Paper~1 that there is no evidence for a significantly 
reduced degree of completeness in the HES sample in this redshift range.

\noindent $\mathbf{0.7 < z < 1.25}$: \ 
The redshift distribution seems to have a local maximum around
$z\simeq 1.1$, but
this is unlikely to be a positive selection bias: There are no strong
emission lines visible (only C\,{\sc iii}] $\lambda$1909 which is
often just barely detected), and neither the colour
dependence of $z$ nor the $K$ correction have discernible features at this
redshift. The most probable interpretation is that of a statistical 
fluctuation.

\noindent $\mathbf{1.25 < z < 2.0}$: \ 
The bright end of the QLF in this shell is dominated by the exceptionally
luminous QSO HE~0515$-$4415. In this redshift range,
the total number of QSOs predicted by the B91 model is consistent with
what is observed in the HES; however, there is a tendency that 
the number of fainter objects is predicted too high. This trend is
continued, even stronger, 
in the LBQS as noted by Hewett et al.\ (\cite{hewett*:93:EBOQ}), 
and it is also visible in the HBQS 
(La Franca \& Cristiani \cite{lafr+cris:97:HBQS}), and
it is therefore likely that not incompleteness of the data, but an
overprediction of the PLE model causes the discrepancy.

\noindent $\mathbf{2.0 < z < 2.9}$: \ 
The high-redshift range of the present HES sample is still only
sparsely populated. Up to $z\simeq 2.6$, the HES selection criteria are
predominantly sensitive to continuum properties. Although Ly$\alpha$ is
generally detected as a strong emission line from $z\ga 1.8$ onward, the broad
range of colour criteria employed in the HES ensures that \emph{all} of
the $z<2.6$ QSOs in the sample are in fact colour-selected (cf.\ Paper~1), 
thus avoiding the strong biases associated with emission line
detection. The expected quasar colours as functions of $z$ show
a strong maximum of UV excess around $z\simeq 2.1$ -- therefore the 
decrease of numbers when passing the $z=2$ mark cannot be related to
selection effects.

At $\mathbf{z>2.6}$, the QSOs were generally found by
feature-detection algorithms,
depending on the presence of a strong L$\alpha$ emission and/or a 
pronounced spectral break at the onset of the Lyman forest. 
Both features are common but not \emph{necessarily} strong,
making it hard to quantitatively assess the
degree of (in)completeness. Another difficulty is the increasingly
ill-defined $K$ correction, with intergalactic Ly$\alpha$ forest
absorption adding to the uncertainty of spectral energy distribution.
The number of objects in the sample should be,
at any rate, considered as a \emph{lower limit} to the number
expected for a `perfect' sample limited only by $B_J$ magnitude.

\begin{figure}[tb]
\epsfxsize=8.8cm
\epsfclipon
\epsfbox[75 92 412 358]{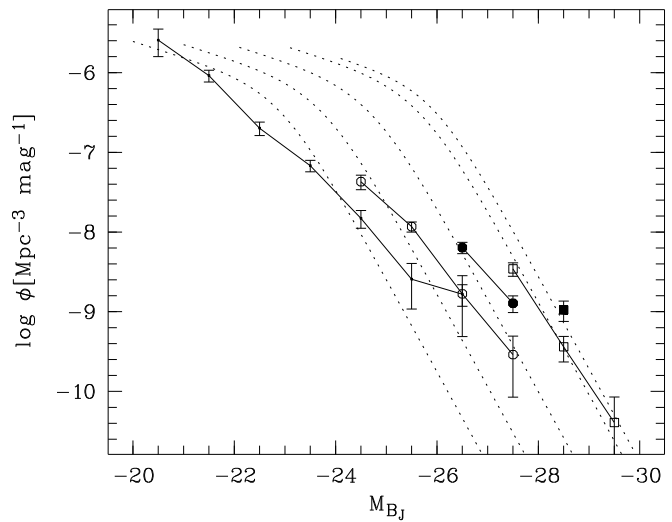} 
\caption[]{Evolution of the differential QLF 
in bins of 1\,mag with redshift. The redshift shells
are the same as in Fig.\ \ref{fig:cumlf_p}, from left
to right. The dotted line shows the prediction of the
Boyle et al.\ (\cite{boyle*:91:QE}) PLE model.}
\label{fig:difflf_all}
\end{figure}

In Fig.\ \ref{fig:difflf_all} all data have been combined into
one frame, presented in the traditional binned differential form.
The HES provides
substantial improvement over previous surveys in covering the bright 
end of the QLF at all redshifts $z\la 3$, plus a full construction 
of the important \emph{local} luminosity function. 
For comparison, the dotted line in Fig.\ \ref{fig:difflf_all} shows the
differential QLF, integrated over the appropriate redshift ranges,
from the Boyle et al.\ (\cite{boyle*:91:QE}) PLE model.

\section{Constraints on parametric models}

\subsection{Pure luminosity evolution}

By combining the $\sim 400$ QSOs from the AAT multifibre survey with a
set of other, smaller quasar samples -- in particular the BQS at the
bright end --, Boyle et al.\ (\cite{boyle*:88:EOQ}) suggested that the
quasar luminosity function can be approximated as a (logarithmically)
shape-invariant double power law, merely shifted in luminosity by an
offset governed through a redshift-independent evolution parameter 
$k_L \approx 3$--3.5. 
In a later analysis, Boyle et al.\ (\cite{boyle*:91:QE}; hereafter B91)
substantiated this claim and extended the PLE model up to $z=2.9$,
albeit with the introduction of an additional parameter
$z_{\mathrm{max}}$ above which the comoving space densities were kept
constant. 

Comparing the prediction of the specific parametric model
of B91 with the empirical LF estimates based on the new HES sample
(Fig.\ \ref{fig:difflf_all}) reveals
substantial discrepancies at most redshifts:
\begin{itemize}
\item At low and intermediate $z$, there is a significant
excess of high-luminosity QSOs over the PLE prediction.
\item At low redshifts, there are less intermediate-luminosity
QSOs than demanded by PLE.
\item The characteristic `break'
observed at high $z$ does not appear in the local ($z<0.3$) QLF.
\item The QLF is much steeper at high redshift. 
\item There is evidence for continued evolution after $z>2$.
\end{itemize}

Most of these discrepancies, in particular the feature of a relatively
flat low-redshift QLF, were already detected in previous surveys
(Hewett et al.\ \cite{hewett*:93:EBOQ}; 
Miller et al.\ \cite{miller*:93:EQP})
and are clearly not artefacts of the HES selection procedure;
the HES confirms these findings with high significance.
Because the actual low-$z$ QLF is flatter than the PLE prediction,
discrepancies show up most prominently at the highest luminosities.
With its large effective area and the luminosity range probed,
the HES provides considerably improved statistical coverage
to test the validity of the PLE parametrisation.
A two-dimensional Kolmogorov-Smirnov test 
(Peacock \cite{peacock:83:2dGF}) applied to the joint distribution 
of ($z$, $M$) pairs yields a formal acceptance probability 
of $P= 10^{-7}$ for the redshift range $0.3<z<2.2$, and 
$P = 10^{-16}$ for the whole definition interval of the HES,
$0 < z < 3.2$, so the B91 model is rejected with very high
confidence. 

\begin{figure}[tb]
\epsfxsize=8.8cm
\epsfclipon
\epsfbox[42 34 454 357]{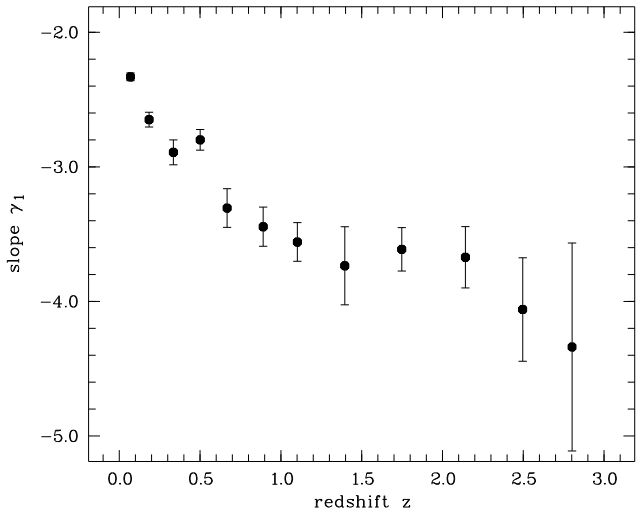}
\caption[]{Variation of the bright-end slope $\gamma_1$ with redshift.}
\label{fig:slopes}
\end{figure}

An essential feature of all PLE models is the fact that 
the slope at the bright end of the QLF, $\gamma_1$, is 
necessarily constant. Fig.\ \ref{fig:difflf_all} shows
that except for the $z<0.3$ shell, the HES objects are
always located on the steep part of the QLF, well beyond
the ubiquitous `break'; this will be so for all realistic 
PLE models. A good test for the validity of PLE models
in general can be made by monitoring the bright-end slope 
of the QLF as a function of $z$. To this purpose we have 
fitted straight lines to the cumulative logarithmic LF 
within each redshift shell (i.e., assuming a power law).
The result is shown in Fig.\ \ref{fig:slopes}, where 
$\gamma_1(z)$ is resolved into redshift bins of 
$\Delta\log (1+z) = 0.05$. Although Malmquist bias and 
evolution within the shells tend to make the measured slope 
at given $z$ somewhat flatter than the \emph{intrinsic} one, 
the bins are small enough to ensure that this bias is almost 
negligible.

The trend for the bright-end slope to become flatter with 
decreasing redshift is very obvious, confirming the earlier
conjecture by Goldschmidt \& Miller (\cite{gold+mill:98:QE}),
with significantly improved redshift resolution.
Fig.\ \ref{fig:slopes} also shows that this trend 
is confined to the redshift regime $z \la 1$, 
while for $z>1$ the slope is consistent with being constant
at $\gamma_1 \simeq -3.7$, approximately equal to the value 
of $\gamma_1$ in most PLE fits (B91; La Franca \& Cristiani
\cite{lafr+cris:97:HBQS}). 
Fig.\ \ref{fig:slopes} may be taken to furthermore suggest 
that the QLF steepens again after $z>2$, but that
is not statistically significant. 

This test demonstrates that the basic assumption of all PLE models --
a redshift-invariant QLF shape -- is in conflict with observations,
independently of the details of the parametrisation.
The bright end of the quasar luminosity function undergoes
a gradual transformation, becoming flatter with cosmic time.

\subsection{Luminosity-dependent luminosity evolution}

In a recent attempt that includes several sets of new survey material
come available in the 1990s, La Franca \& Cristiani
(\cite{lafr+cris:97:HBQS})
modelled the QLF with a `luminosity-dependent \emph{luminosity} evolution' 
(LDLE) scheme, by equipping the PLE formula of Boyle et al.\ with 
an additional term
to produce a flatter QLF at low redshift while maintaining the PLE
description for higher redshifts. We have subjected their model
(including the correction from 
La Franca \& Cristiani \cite{lafr+cris:98:HBQS})
to similar tests as the Boyle et al.\ model, with the
following results:
(1) There is very little difference in performance between LDLE and PLE 
for $z\ga 1$.
(2) The bright end slope $\gamma_1$ decreases (by absolute value) 
towards lower redshifts as demanded by the observations, but the
number of luminous QSOs is \emph{over}predicted.
(3) Below $z \simeq 0.4$ the LDLE formalism assigns
strongly \emph{negative} evolution to the most luminous QSOs,
producing an almost flat ($\gamma_1 = -1$) local luminosity function,
which is of course incompatible with all available data.
In conclusion, the LDLE scheme as proposed by La Franca \& Cristiani
does not resolve the shortcomings of PLE, and it is completely
inadequate for the low-redshift domain.

\subsection{Pure density evolution}

Although pure density evolution (PDE) is the oldest and intuitively
simplest form to parameterise an evolving luminosity function, 
it has been ruled out by several independent investigations.
The argument is usually based on the counts of faint
QSOs at high redshifts, which are observed to be much rarer than
predicted by PDE. Even if faint QSO surveys are assumed to be
substantially incomplete, a stringent limit is set by the 
measured intensity of the diffuse extragalactic soft X-ray background 
(Marshall et al.\ \cite{marshall*:83:ACQS}). 
In a preliminary analysis of a composite ROSAT-selected AGN sample,
Hasinger (\cite{hasinger:98:XLF}) recently challenged the unequivocal
rejection of the PDE concept, claiming that simple density evolution 
in fact provided a much better fit to the data than PLE. This claim
has meanwhile been qualified (Miyaji et al.\ \cite{miyaji*:98:ERLF};
see below), but the performance of PDE with these new samples 
seems indeed surprisingly good.

While a detailed discussion of the similarities and differences
between optically and X-ray selected samples is clearly outside the
scope of this paper, it may be worthwhile to investigate the merits
of PDE for the bright HES sample. Fitting a one-parameter PDE
model to the data is simple: We use the 
modified variable $V'/V'_{\mathrm{max}}$ defined as the
density-weighted volume integrals, $V' = \int \varrho(z)\,dV$,
where $\varrho (z)$ is the density evolution function.
Adopting the customary form for $\varrho (z) = \varrho(0)\,(1+z)^{k_D}$,
$k_D$ is varied until the sample average of $V'/V'_{\mathrm{max}}$
reaches 1/2. Based on the new $K$ correction and using the full
HES sample within $0<z<3.2$, this yields $k_D = 5.35$, thus a much 
smaller value than obtained in earlier fits based on fainter samples,
but very similar to the X-ray value of Hasinger (\cite{hasinger:98:XLF}).
The distribution of $V'/V'_{\mathrm{max}}$ is also
consistent with being uniform within [0,1]. 
However, this is only a necessary and not a sufficient condition;
the two-dimensional KS statistic yields
a formal acceptance probability of $P=10^{-17}$.
For the first time, PDE can be significantly rejected 
from a bright QSO sample alone,
\emph{without any reference to faint QSO counts}.
The reason is, of course, again the steepening 
of the bright QLF part towards higher redshift, leading to
incompatible QLF shapes at low and high $z$. Note that although the
$z\simeq 0$ and the $z\ga 2$ QLF sections as seen in the HES do not
overlap in luminosity, there is sufficient dynamic range at
intermediate redshifts to monitor the steepening slope at 
given $M_{B_J}$.

\subsection{Luminosity-dependent density evolution}

A luminosity function changing with cosmic time can be always be 
described by the general concept of luminosity-dependent 
density evolution (LDDE).
Concentrating on the more luminous part of the QSO
population, Schmidt \& Green (\cite{schm+gree:83:QEBQS}) proposed a
specific LDDE scheme in which the more
luminous QSOs undergo stronger evolution. (Note that this is
not necessarily in conflict with the principles of PLE, which could be
translated into an equivalent LDDE formalism.) However, in the Schmidt
\& Green picture, the QLF becomes successively steeper at smaller
redshift, opposite to what is observed in the new samples.

A related approach was followed by Miyaji et al.\
(\cite{miyaji*:98:ERLF}), accommodating the proposed density evolution 
of soft X-ray selected AGN (Hasinger \cite{hasinger:98:XLF})
with the constraints set by the X-ray background. 
In their formulation, the density evolution 
index $k_D (L)$ is smallest on the low-luminosity tail 
of the QLF; above a certain luminosity threshold, 
the index remains constant. The latter feature implies that 
directly transfering this model to the evolution of optically 
selected QSOs would result in the same conflicts as for the case 
of PDE discussed in the preceeding subsection. 
The specific merits of the Miyaji et al.\ LDDE model are in the 
low-luminosity regime, where the HES yields \emph{no} constraints 
on the evolution rates, apart from the local luminosity function.

While a single survey such as the HES provides insufficient coverage
of the Hubble diagram to allow new parametric models to be 
developed, it seems safe to conclude from the present data 
that only a rather complex luminosity-dependent evolution scheme 
will be capable to reproduce all observed features, making quasar 
evolution more complicated than previously thought.

\section{The local luminosity function  \label{sec:lqlf} }

The basic results of our earlier analysis of the local QLF 
(K\"ohler et al.\ \cite{koehler*:97:LQLF}; hereafter K97) were:
(1) The number of high-luminosity low-redshift QSOs has previously been
underestimated. 
(2) The local QLF does not display any evidence of a break or
significant change of slope at absolute magnitudes $M_{B_J}\la -20$.
The new HES sample improves the K97 data by a factor of 6 in covered
area and by a factor of 8 in sample size.
Both above mentioned results are confirmed in the present paper,
although the discrepancy in number counts is less extreme than
found in the K97 sample. However, considering the error bars,
the results of the two analyses are fully compatible.

A new feature needing corroboration is
the apparent flattening of the local QLF at luminosities 
fainter than $M_{B_J}\simeq -20$,
apparent in Fig.\ \ref{fig:cumlf_p}. Note that 
it is based on only 3 sources with $M_{B_J} > -20$,
thus the statistical significance of the feature is very low.
We are currently investigating this part of the local QLF in more 
detail, including a dedicated treatment of host galaxy influences
and including the transition to low-level AGN in nearby galaxies.
For the present discussion we concentrate on objects with
$M_{B_J} < -20$. Even for these, host galaxy contributions
to standard isophotal magnitudes would be non-negligible.
To avoid this effect, the HES magnitudes were measured over
effective apertures of the size of the seeing disk, so the absolute
magnitudes do not correspond to \emph{total}, but effectively to
\emph{nuclear} luminosities; even large and bright host galaxies 
do not contribute with more than their seeing-convolved central surface
brightness to the measured flux. Thus, while the measured magnitudes
are not individually corrected for host galaxy contributions as it was
the case in K97, they are nevertheless dominated by the active
nuclei, not by the galaxy hosts. 

The accurate determination of the local luminosity function of QSOs 
and Seyfert nuclei is an important, but also a difficult task.
In addition to the challenge of generating unbiased samples, 
there are several problems to be overcome: Disentangling the
contributions of nuclei and hosts to the measured luminosity
is a problem not only in the optical domain, but also at X-ray
wavelengths (Lehmann et al.\ \cite{lehmann*:98}). A further
complication arises from the probable bifurcation of the
AGN population into a Seyfert~1 and a Seyfert~2 branch.
These points need to be addressed before a meaningful physical 
interpretation of the local QLF can be made.

\section{The most luminous QSOs at high redshift}

\subsection{Evolution of space densities}

The substantial uncertaintes in the present knowledge of quasar
evolution at redshifts $z>2$ are mostly related to the increasing
difficulties of obtaining sizeable well-defined flux-limited samples.
The traditional UV excess method based on $U-B$ colours breaks 
down at $z=2.2$ where the Ly$\alpha$ emission line is redshifted
into the $B$ band; multicolour techniques are capable of reaching
higher $z$, but with the penalty of a complicated redshift-dependent 
selection function.
Photographic objective prism surveys such as the LBQS and the HES, 
on the other hand, perform quite well up to $z\simeq 3$, 
but become rapidly incomplete towards $z=3.4$ where Ly$\alpha$ 
moves out of the detector bandpass (see also Sect.\ \ref{sec:lf}).
While the usefulness of the HES sample for space density estimation
is therefore limited to $z\la 3$, this is just the redshift region
where most previous workers located the maximum of quasar activity. 
Furthermore, multicolour surveys show greatly reduced performance 
in this $z$ range, including the forthcoming SDSS (Fan \cite{fan:99:SSO}). 

While at low redshift the sampled luminosity range of a flux-limited
survey depends sensitively on $z$, this dependency is much weaker
towards higher redshift, and here it is possible to trace
quasar space densities \emph{for given luminosity} over a wide
redshift range. The HES provides appropriate data material to 
apply this to the rare species of very high luminosity QSOs,
following their evolution up to $z\simeq 3$.

\begin{figure}[tb]
\epsfxsize=8.8cm
\epsfclipon
\epsfbox[42 34 454 357]{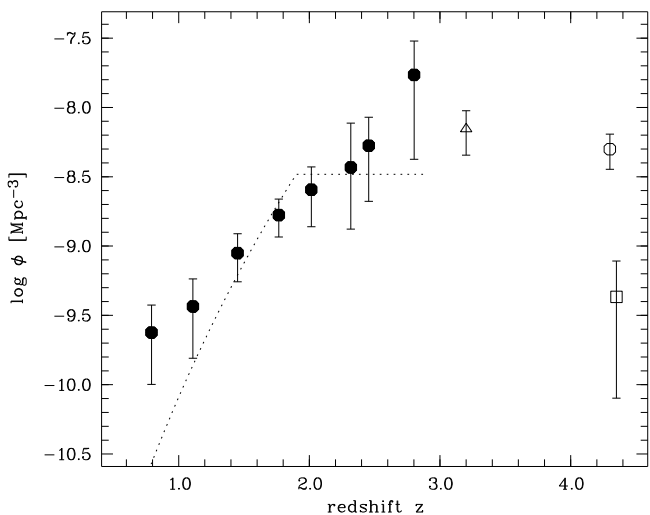} 
\caption[]{Evolution of the integrated space density for QSOs with
$M_B < -28$. The filled symbols show the values computed from the HES,
the dotted line delineates the prediction of the B91 PLE model.
Open symbold indicate the available higher redshift measurements for
this luminosity regime
(triangle: from Hewett et al.\ \cite{hewett*:93:EBOQ}; 
circle: Irwin et al.\ \cite{irwin*:91:HZQ};
square: Kennefick et al.\ \cite{kennefick*:95:LFQ}).}
\label{fig:highz}
\end{figure}

Fig.\ \ref{fig:highz} shows the evolution of integrated 
comoving space density for $M_{B_J} < -28$ 
over several disjoint redshift bins.
Beyond $z>2.5$, even the faintest HES quasars are more luminous than
this. The last plotted value is therefore based on a 
linear fit to the (logarithmic) cumulative luminosity function 
in the corresponding bin, extrapolated to $M_{B_J} = -28$. 
This increases the uncertainty, but does not bias the estimate.
Recall also that QSO selection gets increasingly difficult close 
to the HES high-redshift limit, so that the true space density
for this last point will be even higher.

The steady increase in space densities, up to the highest redshifts
sampled, is remarkable. In particular, there is no indication for a 
turnover or only a slowing down of evolution.
Unless the space density as a function of redshift shows an 
extremely narrow peak just below $z=3$ -- which would still be
consistent with our data, considering the error bars --, we conclude
that the maximum space density has almost certainly not yet been 
reached by our survey and must be located well beyond $z=3$. 
This is in marked contrast to the general notion that the maximum
of quasar activity occurs at $z<3$ 
(e.g., Schmidt et al.\ \cite{schmidt*:95:ELFQ}; 
Shaver et al.\ \cite{shaver*:96:SDQ}).
Two possible interpretations are conceivable:

(1) The maximum of quasar activity might depend on luminosity, in the
sense that high-luminosity objects have a maximum shifted towards
higher redshifts. 
Only three published surveys can provide meaningful additional datapoints 
for Fig.\ \ref{fig:highz}: The estimate for $3<z<3.4$ 
from the LBQS (Hewett et al.\ \cite{hewett*:93:EBOQ})
which is almost certainly only a lower limit; the $z>4$ survey
from Irwin et al.\ (\cite{irwin*:91:HZQ}), and the similar sample
from Kennefick et al.\ (\cite{kennefick*:95:LFQ}). None is in
direct conflict with the HES data, but the high space density
claimed by Irwin et al.\ (\cite{irwin*:91:HZQ}) seems to be
joining particularly well with the HES. However, the origin
of the discrepancy between the results of Irwin et al.\ 
and Kennefick et al.\ -- based on partly the same objects -- 
is not sufficiently understood. Kennefick et al.\ suggested that 
different adopted $K$ corrections may be responsible, but 
this has not been confirmed so far.

Additional support for a stronger evolution of high-luminosity QSOs
comes from the analysis of Warren et al.\ (\cite{warren*:94:MSQ3}): 
although their survey does not reach the luminosity range in question, 
their adopted luminosity evolution model in which only the bright part 
evolves can be extrapolated to evolution rates around $z\simeq 2$--3 
that are quite compatible with the HES data.
 
If it should prove correct that high-luminosity QSOs continue to 
show positive evolution where the space densities of their
lower luminosity counterparts already turn over, this
has substantial consequences for cosmogony and galaxy formation
scenarios. High luminosity presumably implies host galaxy
mass (in analogy to the confirmed such relation at low redshifts,
cf.\ McLeod \& Rieke \cite{mcle+riek:95}). This would impose
significant constraints on theories for the formation of 
very massive structures in the early universe.

(2) Incompleteness of fainter QSO surveys provides an obvious 
alternative explanation. Several recent deep surveys found
high-redshift QSOs in larger numbers than anticipated
(Miyaji et al.\ \cite{miyaji*:98:ERLF}; 
Wolf et al.\ \cite{wolf*:99:HRQ}),
although intercomparisons are difficult as each survey samples
a different luminosity regime, and the statistics are still
very poor.

\subsection{Gravitational lensing}

Statistical analyses based on optical quasar surveys can
suffer from biases quite similar to those present in low-redshift 
QSO host galaxy studies: Systematic rejection of non-pointlike
sources, a property of most surveys, invariably leads to missing
large-separation lenses. The HES avoids this selection bias,
and since since it samples the region of the QLF where 
`magnification bias' (Turner \cite{turner:80:GLQ})
is expected to be most effective,
it is well suited as a testing ground for lensing statistics.
While a full investigation of this subject is beyond the scope
of this paper, we can make a few qualitative inferences on the effects
of lensing onto the observed luminosity function.

The principal result of `magnification bias' is to create a flatter
observed QLF, compared to the intrinsic one. It can be shown that
the effect is noticeable only when the QLF slope is steep
(e.g., Schneider et al.\ \cite{schneider*:92:GL}). 
As the optical depth for gravitational lensing 
increases with $z$, the magnification bias is expected to be 
stronger for sources at higher redshift. 
Under the assumptions of an intrinsically constant QLF slope
and magnification bias being a dominant effect, 
we should observe a bright end of the QLF
that flattens with increasing redshift, 
opposite to what is actually observed.
This does not exclude that there might be \emph{some} influence 
of lensing on the observed luminosity function, 
but the observations constrain magnification bias 
to be of very minor importance for the QLF 
as a whole up to $z\simeq 2$.
This is consistent with the fact that only  
$\sim$ 1 out of 100 QSOs with $M_V <-28$ shows
discernible image splitting due to lensing
(Surdej et al.\ (\cite{surdej*:93:GLS}).

\section{Conclusions}

The analysis of a new sample of very bright QSOs has shown
that quasar evolution is more complex than generally assumed.
The Hamburg/ESO survey provides insight into otherwise rarely
sampled regions of the luminosity-redshift plane, enabling 
a direct assessment of the local luminosity function, 
and to probe the evolution of the bright part of the QLF
up to $z\simeq 3$. While the high-redshift QLF has a very steep
high-luminosity tail, the low-redshift QLF approaches towards 
a simple power-law form with rather flat slope.
No simple `pure' evolution scheme based on a shape-invariant QLF is capable
of reproducing this transformation. Future analyses of large composite
samples covering a wide range of luminosities are required to show if 
appropriate parametric descriptions can be found that are consistent
with this behaviour.

\end{document}